\UseRawInputEncoding

\documentclass[journal]{IEEEtran}
\ifCLASSINFOpdf
  \usepackage[pdftex]{graphicx}
  \graphicspath{ {figs} }
  \DeclareGraphicsExtensions{.pdf,.jpeg,.png}
\else
\fi
\usepackage[table,xcdraw]{xcolor}
\usepackage{tabularx}
\usepackage{multirow}

\usepackage[]{changes}
\definechangesauthor[name={liangxiu han},color= blue]{LHAN}
\usepackage{csquotes}
\usepackage{todonotes}
\usepackage[switch]{lineno}

\begin{document}
%
\title{An Explainable 3D Residual Self-Attention Deep Neural Network For Joint Atrophy Localization and Alzheimer's Disease Diagnosis using Structural MRI}

%
%
%

\author{Xin Zhang, Liangxiu Han, Wenyong Zhu, Liang Sun, Daoqiang Zhang

\thanks{ Xin Zhang and Liangxiu Han are with the Department of Computing, and Mathematics, Manchester Metropolitan University, Manchester M15GD, U.K (e-mail: x.zhang@mmu.ac.uk; l.han@mmu.ac.uk)}
\thanks{ Wenyong Zhu, Liang Sun, and Daoqiang Zhang are with the College of Computer Science and Technology, Nanjing University of Aeronautics and Astronautics, Nanjing 210016, China (e-mail: wyzhu@nuaa.edu.cn; sunl@nuaa.edu.cn; dqzhang@nuaa.edu.cn)
}
\thanks {Corresponding authors: L. Han and D. Zhang (e-mail: l.han@mmu.ac.uk; dqzhang@nuaa.edu.cn)}

}

       

%
%

\markboth{Journal of \LaTeX\ Class Files,~Vol.~14, No.~8, August~2015}%
{Shell \MakeLowercase{\textit{et al.}}: Bare Demo of IEEEtran.cls for IEEE Journals}
%



\maketitle

\begin{abstract}
Computer-aided early diagnosis of Alzheimer's disease (AD) and its prodromal form mild cognitive impairment (MCI) based on structure Magnetic Resonance Imaging (sMRI) has provided a cost-effective and objective way for early prevention and treatment of disease progression, leading to improved patient care. 
In this work, we have proposed a novel computer-aided approach for early diagnosis of AD by introducing an explainable 3D Residual Attention Deep Neural Network (3D ResAttNet) for end-to-end learning from sMRI scans.  Different from the existing approaches, the novelty of our approach is three-fold: 1) A Residual Self-Attention Deep Neural Network has been proposed to capture local, global and spatial information of MR images to improve diagnostic performance; 2) An explainable method using Gradient-based Localization Class Activation mapping (Grad-CAM) has been introduced to improve the interpretability of the proposed method; 3) This work has provided a full end-to-end learning solution for automated disease diagnosis. Our proposed 3D ResAttNet method has been evaluated on a large cohort of subjects from real datasets for two changeling classification tasks (i.e. Alzheimer's disease (AD) vs. Normal cohort (NC) and progressive MCI (pMCI) vs. stable MCI (sMCI)). The experimental results show that the proposed approach has a competitive advantage over the state-of-the-art models in terms of accuracy performance and generalizability. 
The explainable mechanism in our approach is able to identify and highlight the contribution of the important brain parts (e.g., hippocampus, lateral ventricle and most parts of the cortex) for transparent decisions. 

\end{abstract}


%
\IEEEpeerreviewmaketitle

\section{Introduction}
%
%
%
%
\IEEEPARstart{A}{lzheimer's}  disease (AD) is the most common cause of dementia among the old people, which is irreversible and progressive neurodegenerative brain disease. It contributes to 60-70\% of dementia cases and affects over 30 million individuals\cite{who}. With the exponentially growing aging population across the globe, the prevalent increased cases of Alzheimer's disease (AD) have presented unprecedented pressures on public healthcare service. There is currently no cure for AD. However, the progress could be slowed through medicine and optimized physical cognition and activity.  Therefore, accurate and timely diagnosis of Alzheimer's disease (AD) and its early form mild cognitive impairment (MCI) is essential for optimal management and improved patient care \cite{ahmed_alzheimers_2015}. Clinically, Structural Magnetic Resonance Imaging (sMRI) has been used for AD diagnosis. The structural MRI measurement is considered as a marker of AD progression, which can help detect the structural abnormalities and track the evolution of brain atrophy\cite{huang_voxel-_2003,frisoni_clinical_2010,rathore_review_2017}. However, the disease identification process is mainly performed manually by specialists, which is time-consuming and expensive. 

\par To solve this problem, much effort has been devoted to developing computer-aided diagnostic systems for automated discrimination of progression of AD (e.g., mild cognitive impairment (MCI) including progressive MCI (pMCI) and stable MCI (sMCI) and normal cohort (NC)) from sMRI scans based on voxel-wise global features, or predetermined regional features or combination of both \cite{allioui_utilization_2020,hosseini-asl_alzheimers_2016,payan_predicting_2015, vu_multimodal_2017}. The volumetric or voxel-based approaches extract global features for detecting the structure changes and identifying voxel-wise disease associated microstructures for AD diagnosis \cite{baron_vivo_2001, ashburner_why_2001,ashburner2000voxel,mechelli_voxel-based_2005}. The Tensor-based morphology (TBM) diagnostic approach is a voxel-wise optimization approach, which can recognize local structural changes through mapping orders of local tissue volume loss or income over time to understand the neurodegenerative or neurodevelopment processes for AD diagnosis\cite{studholme_deformation-based_2006}. In \cite{kloppel_automatic_2008}, the gray matter voxels were selected as features and used to trained a machine learning model for AD vs. NC classification.  

\par Since some specific brain regions such as hippocampal region of interest (ROI) are strongly correlated to the disease, several existing works focused on some predetermined ROIs guided by prior biological knowledge and extracted regional features for AD diagnosis \cite{ahmed_classification_2015,magnin_support_2009,gerardin_multidimensional_2009, gutman_disease_2009, planche_hippocampal_2017}. For instance, Magnin \cite{magnin_support_2009} and Zhang \cite{zhang_multimodal_2011} applied Support Vector Machine (SVM) to learn regional features for AD diagnosis by splitting the brain into some non-overlapping areas.

\par Recently, deep neural networks have shown successful for various computer vision tasks \cite{he2016deep,lecun2015deep}. A few deep learning methods have been proposed for AD diagnosis with sMRI scans and achieved better performance than the classical machine learning-based methods. These methods focused on learning either regional features from prior Knowledge regions (e.g., hippocampus \cite{mu_adult_2011,gutman_disease_2009}, cortical \cite{billones_demnet:_2016}), global features \cite{hosseini-asl_alzheimers_2016} or combination of both \cite{lian_hierarchical_2018}). Lian et al. proposed a hybrid deep learning approach using convolutional neural networks (CNNs) to learn combined features at multiscale \cite{lian_hierarchical_2018}. Hosseini-Asl et al. predicted AD with a 3D CNNs based on the pretrained 3D convolutional autoencoder model to capture anatomical shape variations from sMRI \cite{hosseini-asl_alzheimers_2016}.   

\par Despite the existing encouraging work, it suffers several limitations. Firstly, extracting global features using voxel-based approaches involve processing high-dimensional 3D data, which is computationally intensive. Secondly, regional-based features focusing on certain brain regions of interest (e.g., the cortical thickness and hippocampus shape) may neglect possible pathological locations in the brain and fail to obtain global structural information for accurate AD diagnosis. Moreover, these methods require domain-specific prior-knowledge and multi-stage training. Thus, it is hard to provide an end-to-end solution for automatic disease diagnosis. Thirdly, the existing methods \cite{lian_hierarchical_2018,hosseini-asl_alzheimers_2016} used combined features or global features to improve disease diagnostic performance based on deep learning approaches. However, the use of hybrid loss functions for each layer with the same shared weight may lead to difficulty in training and reproduction. Finally, most of existing deep learning-based approaches for AD diagnosis lack transparency in terms of model explanation due to the nature of black-box learning.  
\par To overcome the aforementioned limitations, this work proposes a novel computer-aided approach for early diagnosis of AD from sMRI by developing an explainable 3D Residual Attention Deep Neural Network (3D ResAttNet) for end-to-end learning from sMRI. Different from the existing approaches, our contributions lie in:
\begin{enumerate}
  \item A Residual Attention Deep Neural Network has been designed and implemented, allowing for capturing local, global and spatial information to improve diagnostic performance;
  \item An explainable Gradient-based Localization Class Activation mapping (Grad-CAM) has been introduced, enabling visual explanation and interpretation of model predictions;
  \item The proposed work has provided a full end-to-end solution for automated disease diagnosis.
\end{enumerate}
\par The rest of this paper is organized as follows: Section 2 presents related work; Section 3 details the proposed method; In Section 4, the experimental evaluation is described; Section 5 concludes the proposed work and highlights the future work.

\section{Related work}
\subsection{Computer-aided AD diagnosis}
\par Computer-aided diagnosis of AD treatment has a long history, with the aim of extracting useful features for automatic classification. According to the feature extracted method, it can be broadly divided into three categories: 1) Global feature-based approaches (Voxel-based approaches); 2) Regional feature-based approaches; 3) Combination of both global and regional based approaches.  

\par The early works on AD diagnosis mainly focused on the extracted global features from the whole MR image. The volumetric-based approach using voxel intensity features has been widely used for AD classification. Ashburner et al. \cite{ashburner2000voxel} introduced a voxel-based morphometry (VBM) method, which used voxel-wise comparison on the smoothed gray-matter images. It showed the difference between white and gray voxels in local concentrations compared with the normal cohort (NC) brains. Based on the voxel-wise features, Kl\"oppel et al. trained a support vector machine (SVM) model to diagnosis AD from sMRI \cite{kloppel_automatic_2008}. Hinrichs et al. \cite{hinrichs2009spatially} also employed the gray-matter density to extract voxel-wise features, then a linear programming boosting method was trained to classify AD with sMRI images. However, some limitations include 1) computationally intensive and over-fitting due to high dimensionality of features with the relatively small number of images for model training; 2) neglecting the regional information that has been proven important to AD diagnosis.
\par The second category is regional feature-based methods. The majority of the works in this category mainly relied on prior knowledge to determine ROIs. Several existing works in the literature extracted features from the predetermined ROIs based on biological prior knowledge on the shrinkage of cerebral cortices and hippocampi, the enlargement of ventricles, and the change of regional glucose uptake \cite{gerardin_multidimensional_2009, gutman_disease_2009,planche_hippocampal_2017}. Magnin \cite{magnin_support_2009} and Zhang \cite{zhang_multimodal_2011} extracted regional features by splitting the whole brain into smaller regions to train the machine learning classifiers for AD diagnosis. 
The work in \cite{ahmed_classification_2015} used Gauss-Laguerre Harmonic Functions (GL-CHFs) 
and SURF \cite{bay_surf:_2006} descriptors to extract local features from sMRI scans in hippocampus and posterior cingulate cortex (PCC) structures of the brain. Fan \cite{fan_compare:_2006} partitioned the sMRI images into an adaptive set of brain areas based on the watershed algorithm, and then extracted the regional volumetric features to train a SVM-based AD classification model. 
However, these aforementioned methods are based on empirical regions, which might neglect possible pathological locations in the whole brain.  Moreover, the features extracted from ROIs may not be able to reflect the subtle changes involved in the brain \cite{suk_hierarchical_2014}.  
\par In order to address these limitations, a hierarchical method was introduced by combination of global and regional features. Lian et al. divided sMRI images into small 3D patches and extracted features, and then combined the features hierarchically \cite{lian_hierarchical_2018}. Suk et al. also proposed a systematic method for a joint feature representation from the paired patches of sMRI images using the patch-based approach\cite{suk_hierarchical_2014}. These patch-based methods have been proven to efficiently deal with the problem of high dimensional features and also the sensitivity to slight changes. However, these models always require multi-stage training, which are not an end-to-end solution. 


\par Recently, deep learning has achieved a remarkable success in the field of Computer Vision, which has also become a popular and useful method for medical image analysis including Alzheimer's disease (AD) diagnosis based on MRI images. 
The convolutional neural network (CNN) \cite{krizhevsky2012imagenet} \cite{simonyan2014very} has been proven to be suitable for grid-like data such as RGB images and MRI images.  
Billones et al. proposed a modified 16-layered VGG network to AD classification with sMRI images \cite{billones_demnet:_2016}. The method selected 20 central slices of a sMRI image and achieved high accuracy on classification tasks using 900 sMRI images from the ADNI database. Residual network is the most widely used CNNs architecture that won the Imagenet classification competition \cite{he2016deep}. It aims to alleviate the issues with the vanishing/exploding gradients when the network becomes deeper. In ResNet Block, a shortcut connection is added to link the input with the output, thus the Resnet learns the residual of input. Li et al.\cite{li_deep_2017} proposed a deep network with residual blocks for AD diagnosis using 1776 sMRI images from the ADNI database.

\subsection{Explainable deep learning}
\par Due to the nature of black box, one challenge facing in the deep learning models is their explainable capability \cite{sample_computer_2017}. For the AD diagnosis task, most of existing deep learning-based approaches lack transparency with difficulty in explaining why and how a model decision is reached. 
To explain the image classification result by CNNs models, several explainable methods for CNNs have recently been proposed. 

\par Saliency map \cite{simonyan2014very} was firstly used for interpreting CNNs based models, which can highlight and explain which part of image features that contribute the most to the activity of a specific layer in a network or the decision of the network as a whole. 
It computes the gradients of logits based on the backpropagation algorithm and visualizes the feature contributions based on the amount of gradient they receive. 
This Saliency map is suitable for visualization but not good for localization and segmentation due to the noisy results \cite{adebayo2018local}.  Some improved methods based on saliency map \cite{alber2019innvestigate,bojarski2016visualbackprop} have been proposed. For instance, the most widely used method is the guided backpropagation by preventing the backward flow of negative gradients on ReLU activation from the higher layer in the CNNs architecture \cite{springenberg2014striving}. Other optimized visualization methods also include PatternNet and PatternAttribution \cite{kindermans2017learning}, Layer-Wise Relevance Propagation (LRP) \cite{bach2015pixel} and DeepTaylor \cite{montavon2017explaining}.

\par Class Activation mapping (CAM) is another explainable method for CNNs. In the CAM method, the top fully connected layers was replaced by convolutional layer to maintain the object positions and can find the spatial distribution of distinguished regions for predict category \cite{zhou2016learning}.

\par  The CAM requires retraining the model since it changes the model architecture. However, to address this issue, Grad-CAM has been proposed as a generalization of the CAM method \cite{selvaraju2017grad}, which keeps the origin classification architecture and calculates the weight by pooling the gradient. This method has been widely used to explain the CNN classification models. However, since the Grad-CAM extracts the spatial distribution from the last layer of the feature map with low resolution, this results in smaller size than the input image size. In order to obtain more accurate location information at high resolution, some optimized CAM methods are proposed, such as Adversarial Complementary Learning for Weakly Supervised Object Localization (ACOL) \cite{zhang2018adversarial}, Self-produced Guidance for Weakly-supervised Object Localization (SPG) \cite{zhang2018self1} and guided attention inference networks (GAIN) \cite{li2019guided}. To the best of our knowlege, only a few works presented the explainable methods for deep learning based AD Diagnosis. Montavon et al. and Yange et al. \cite{montavon2017explaining, yang_visual_2018} tried to explain 3D-CNNs by using visual interpretation methods. These methods are able to show how the CNNs made the classification decision. But there is no attempt made yet to explain 3D data classification tasks for diagnosis of MCI.

\section{Method}
\par The aim of this work is to develop an end-to-end deep learning framework to automatically classify discriminative atrophy localization on sMRI image for AD diagnosis, which consists of two levels of classifications: Alzheimer's disease (AD) vs. Normal cohort (NC) and progressive MCI (pMCI) vs. stable MCI (sMCI).

\subsection{3D Explainable Residual Self-Attention Convolutional Neural Network (3D ResAttNet)}
\label{sec:3DResnet}

\par
\begin{figure}[h]
    \centering
    \includegraphics[width=0.45\textwidth]{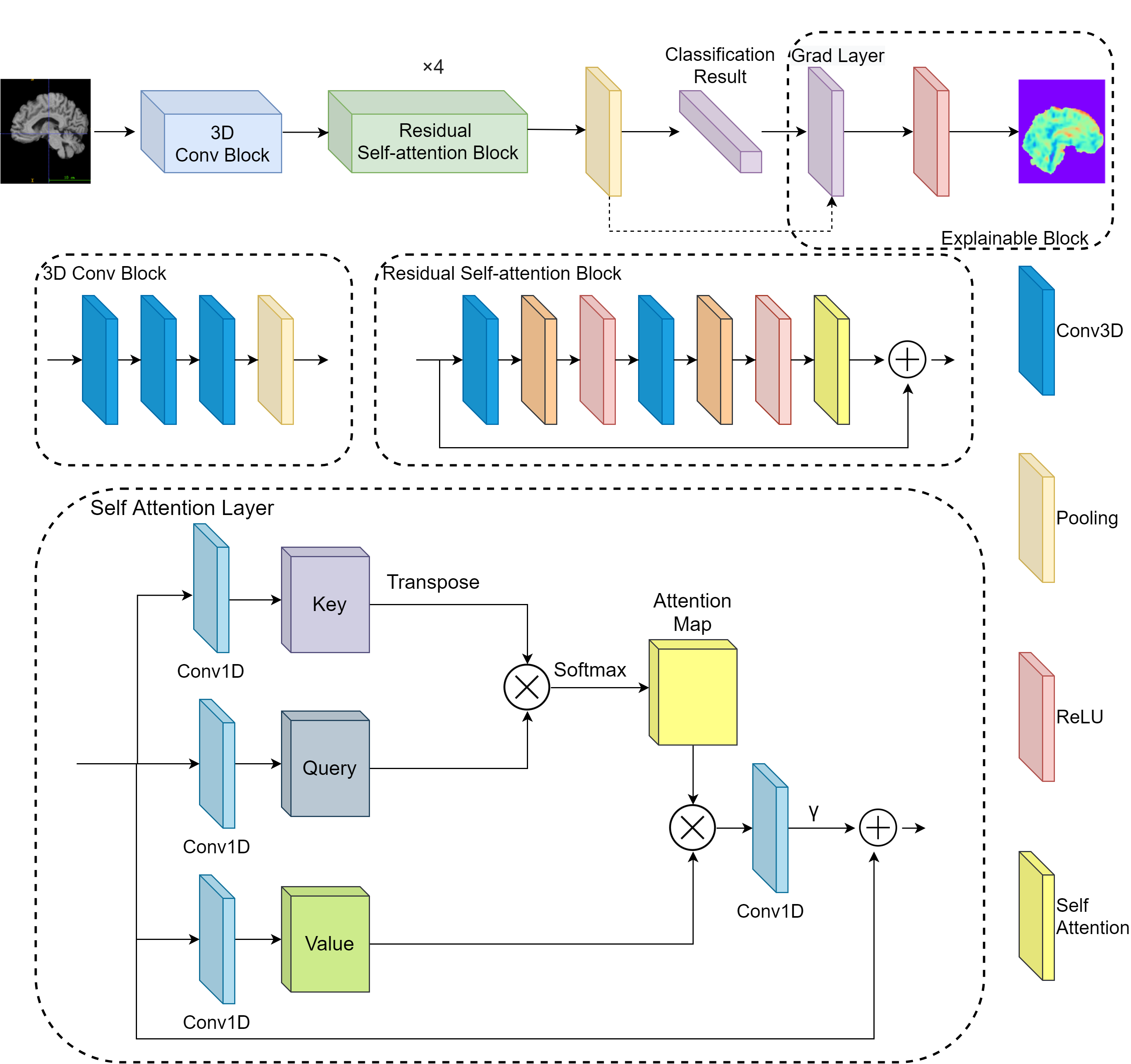}
    \caption{The architecture of 3D residual attention deep Neural Network}
    \label{FIG:1}
\end{figure}

 We have proposed a 3D explainable residual attention network (3D ResAttNet), a deep convolutional neural network that adopts self-attention residual mechanism and explainable gradient-based localization class activation mapping (Grad-CAM) for AD diagnosis. The high-level conceptual framework is shown in {Fig.~\ref{FIG:1}}, which consists of several major building blocks including 3D Conv block, Residual Self-attention block, and Explainable blocks. 
The rationale behind of this architecture design includes:
\begin{enumerate}
  \item The residual mechanism is designed to allow for more efficient training with fewer parameters for performance enhancement when increasing the depth of the network. Existing methods \cite{he2016deep} have shown that residual learning can alleviate the issue of disappearance/exploding gradients when the network becomes deeper. In addition, the residual connection avoids losing global features to ensure the integrity of the original information \cite{wang_residual_2017}.
  \item The self-attention mechanism is added to learn long-range dependencies. Capturing long-range dependencies is important in deep learning. Since the convolutional operator has a local receptive field, the long-distance dependencies can only be captured when repeatedly applying convolutional operations \cite{vaswani2017attention,zhang2018self, wang2018non, buades2005non}, resulting in computational inefficiency.  Hence, it is necessary to add self-attention mechanism to address these issues.
  \item The gradient-based localization class activation mapping (Grad-CAM) is introduced to provide visual explanations of predictions of Alzheimer's disease. 
\end{enumerate}

\subsection{3D CNNs}
 
\par Deep convolutional neural networks provide an effective way to learn multi-level features with multi-layers of convolutional operations in an end-to-end fashion \cite{lecun2015deep}. Essentially, the high-level features are obtained by composing low-level features and the levels of features can be enriched by the number of stacked layers (i.e. depth). We have used 3D CNNs. 3D convolutions apply a 3D filter to the dataset and the filter moves 3 directions x, y, z to calculate the low-level feature representations of the output shape as a 3-dimensional volume space. The stack of three $3\times3\times3$ convolutional layer is used to improve computing efficiency, compared to the widely used $7\times7\times7$ convolutional layer. 

\subsection{Residual Self-attention block (ResAttNet)}
\par In this work, for the first time, we have combined self-attention with residual module to capture both global and local information based on 3D images to avoid information loss. Attention mechanism has been a popular and useful tool in recent year \cite{vaswani2017attention,zhang2018self}, which can learn and focus on the critical areas and suppress non-essential information through a weight matrix in whole image. On the other hand, this may cause global information loss. Therefore, we have added residual connection to address this issue. Residual network was originally designed to solve the issue of disappearance/exploding gradients when a network becomes deeper \cite{he2016deep}. A residual connection is added between the origin input and the processed layer, which allows gradients to propagate more easily through the network.

\subsubsection{Residual network layer} \hfill
\par A residual network can be formulated as follows:
 \begin{equation}
y=x+r(x)
\end{equation}

\par Where $y$ is the output of the residual module, $x$ is the input and $r(x)$ is the residual function. This module includes two Conv3D blocks consisting of $3\times3\times3$ 3D convolution layers, 3D batch normalization and rectified-linear-unit nonlinearity layer (ReLU). 
\subsubsection{Self-attention layer} \hfill
\par As described in Section \ref{sec:3DResnet}, the convolutional operator has a local receptive field and only performs local operations, while the self-attention mechanism can perform non-local operation, capable of capturing long-range dependencies/global information. Therefore, inspired by the work in \cite{buades2005non,wang2018non}, in our work, the self-attention layer is introduced to the end of original residual module $r(x)$ to help the model efficiently capture the global information. A self-attention function can be described as mapping a query, a key, and a value to the input, where those are all vectors. Key and value are the features of the whole sMRI extracted by each convolution block and the query determines which values to focus on for learning process. By using the $1\times1\times1$  convolution filter, the key, query, and value are transformed to vectors. The key, query and value are denoted by $f(x)$, $g(x)$, $h(x)$ as follows:
\begin{equation}
Key:\ f\left(x\right)=W_fx
\end{equation}
\begin{equation}
Query:\ g\left(x\right)=W_gx
\end{equation}
\begin{equation}
Value:\ h\left(x\right)=W_hx
\end{equation}

\par Here $x\in R^{C\times N}$ is the features from the previous layer. C is the number of channels and N is the number of locations of features from the previous layer.  $W_f$, $W_g$ and $W_h$ are all $1\times1\times1$ convolution filters. The self-attention map $(a_{i,j})$ can be calculated as:
\begin{equation}
a_{i,j}=\frac{exp({f\left(x_i\right)}^Tg(x_j))}{\sum_{i=1}^{n}{exp({f\left(x_i\right)}^Tg(x_j))}}
\end{equation}

\par where $a_{i,j}$ indicates the correlative degree of attention between each region $i$ and all other regions. $j$ is the index of an output position. The output of the attention layer is $o=(o_1,o_1\ldots o_j,o_N)\in R^{C\times N}$, where
\begin{equation}
o_j=W_v(\sum_{i=1}^{N}{a_{i,j}h\left(x_i\right)})
\end{equation}
\par In order to keep the same number of channels as the original input and for memory efficiency, a $1\times1\times1$ convolution filter $(W_v)$ is used to reduce the channel number of final outputs.

\subsubsection{Residual Self-attention block (ResAttNet)} \hfil

\par Therefore, the final output of the Residual Attention Block is given by:
\begin{equation}
y=x+r(x)+\gamma o(r(x))    
\end{equation}
\par Where the $o(r(x))$ is the output of self-attention map, $r(x)$ is the output of original output of residual function and x is input feature, the  $\gamma$ is a learnable parameter. We set $\gamma$ as 0 as default to allows the network to first rely on the cues in the local neighborhood. When $\gamma$ increased, the model gradually learns to assign more weight to the non-local evidence.

\subsection{The explainable 3D-CNNs}
\par To understand inside the proposed deep model, the 3D Grad-CAM have been applied to explain the model decision. 
\par We first calculated the gradient of the probabilities of disease areas with respect to the activation of unit $k$ at location $x$, $y$, $z$ in the last convolutional layer of the network. Then, the global average pooling of the gradients $(a_k^c)$ is used to show the importance weights for unit $k$.
\begin{equation}
a_k^c=\frac{1}{Z}\sum_{x}\sum_{y}\sum_{z}\ \frac{\partial y(c)}{\partial A_{x,y,z}^k}
\end{equation}
\par where $Z$ is the number of voxels in the corresponding convolutional layer. Then, we combined the unit weights with the activations, $A_{x,y,z}^k$, to get the heatmap of 3D gradient-weighted class activation mapping. 
\begin{equation}
L_{3D-Grad-CAM}^c=ReLU\left(\sum_{k}{a_k^cA_{x,y,z}^k}\right) 
\end{equation}

\section{Experimental evaluation}

\subsection{Dataset description}
\label{sec:dataset}
\par  Data used in this study are from the ADNI (http://adni.loni.usc.edu), consisting of baseline MRI scans of 1407 subjects from ADNI-1, ADNI-2 and ADNI-3 datasets. These subjects are divided into three classes: AD (Alzheimer's disease), MCI (mild cognitive impairment) and NC (normal control) based on the standard clinic criteria (e.g., Mini-Mental State Examination (MMSE) scores and Clinical Dementia Rating (CDR)). For MCI conversion prediction, MCI subjects are further divided into two classes: pMCI (progressive MCI subjects who had converted to AD within 36 months after baseline visit) and sMCI (stable MCI subjects who were continuously diagnosed as MCI). 
The ADNI-1 consists of 1.5T T1-weighted MR images, which has 835 scans of four classes: 200 of Alzheimer's Disease (AD) patients and 231 of Normal Cohort (NC), 232 sMCI and 172 pMCI. The ADNI-2 dataset consists of 3T T1-weighted MR images, which contains 258 scans of four classes: 108 of AD patients and 150 of NC. The ADNI-3 dataset consists of 3T T1-weighted MR images similar to ADNI-2, which has 314 scans of four classes: 45 of AD patients and 269 of NC.  
The demographic information of subjects is presented in Table~\ref{table:1}.
\par In this work, we have used ADNI-1 for our model construction. ADNI-2 and ADNI-3 dataset have been used for independent cross-validation of model generalizability.


\begin{table}[ht]
\caption{Demographic information in the used dataset.Gender reports are male and female. The age, education years, and mini-mental state examination (MMSE) values are reported.}
\label{table:1}
\centering
\resizebox{0.45\textwidth}{!}{%
\begin{tabular}{cccccc}
\hline
\multirow{2}{*}{Dataset} & \multirow{2}{*}{Group} & Gender              & Age             & Edu            & MMSE           \\ \cline{3-6} 
                         &                        & (Male/Female)       & (Mean $\pm$ Std)      & (Mean $\pm$ Std)     & (Mean $\pm$ Std)     \\ \hline
\multirow{4}{*}{ADNI-1}  & AD                     & 200   ( 103 / 97 )  & 75.62 $\pm$ 7.70  & 14.68 $\pm$ 3.20 & 23.29 $\pm$ 2.04 \\
                         & pMCI                   & 172 ( 106 / 66 )    & 76.34 $\pm$ 7.15    & 15.76 $\pm$ 2.84   & 26.61 $\pm$ 1.70   \\
                         & sMCI                   & 232 ( 154 / 78 )    & 76.47 $\pm$ 7.82    & 15.58 $\pm$ 3.17   & 27.31 $\pm$ 1.79   \\
                         & NC                     & 231   ( 119 / 112 ) & 75.99 $\pm$ 5.00  & 16.06 $\pm$ 2.84 & 29.12 $\pm$ 0.99 \\ \hline
\multirow{2}{*}{ADNI-2}  & AD                     & 108   ( 60 / 48 )   & 74.95 $\pm$ 7.80  & 15.88 $\pm$ 2.66 & 23.03 $\pm$ 2.14 \\
                         & NC                     & 150   ( 73 / 77 )   & 74.84 $\pm$ 6.60  & 16.63 $\pm$ 2.48 & 29.09 $\pm$ 1.19 \\ \hline
\multirow{2}{*}{ADNI-3}  & AD                     & 45   ( 25 / 20 )    & 74.87 $\pm$ 8.701 & 15.98 $\pm$ 2.22 & 22.76 $\pm$ 3.58 \\
                         & NC                     & 269   ( 97 / 172 )  & 70.72 $\pm$ 6.50  & 16.80 $\pm$ 2.25 & 29.09 $\pm$ 1.11 \\ \hline
\end{tabular}%
}
\end{table}

\par As the original dataset is in Neuroimaging Informatics Technology Initiative (NIfTI) format, the preprocessing is needed for spatial distortion correction caused by gradient nonlinearity and B1 field inhomogeneity. This is a standard pipeline process including anterior commissure (AC)-posterior commissure (PC) correction, intensity correction \cite{sled_nonparametric_1998} and skull stripping \cite{wang_robust_2011}. We have used MIPAV(Medical Image Processing, Analysis, and Visualization) application to complete AC-PC correction and use FSL(FMRIB Software Library v6.0) to complete skull stripping. A line align registration strategy (flirt instruction in FSL) is also executed to align every sMRI linearly with the Colin27 template \cite{holmes_enhancement_1998} to delete global linear differences (including global translation, scale, and rotation differences), and also to re-sample all sMRIs to have identical spatial resolution. 


\subsection{Evaluation metrics}
\par We have evaluated two binary classification tasks of AD classification (i.e., AD vs. NC) and MCI conversion prediction (i.e., pMCI vs. sMCI). The classification performance has been evaluated based on four commonly used standard metrics, including classification accuracy (ACC), sensitivity (SEN), specificity (SPE), and Area under the curve (AUC). These metrics are defined as:
\begin{equation}
ACC=\frac{TP+TN}{TP+TN+FP+FN} 
\end{equation}
\begin{equation}
SEN=\frac{TP}{TP+FN}                                               
\end{equation}
\begin{equation}
SPE=\frac{TN}{TN+FP}                                                        
\end{equation}
\par where $TP = True Positive$, $TN= True Negative$, $FP= False Positive$ and $FN=False Negative$. The $AUC$ is calculated based on all possible pairs of $SEN$ and $1-SPE$ obtained by changing the thresholds performed on the classification scores yielded by the trained networks.

\subsection{Experimental evaluation}
\par To evaluate performance and generalizability of our proposed model, we have conducted three types of experiments: 1) Comparison study with state-of-the-art 3D convolutional neural networks; 2) Evaluation on generalizability of the proposed model using two independent datasets (ADNI-2 and ADNI-3); 3) Comparison study with other existing machine learning/deep learning methods for AD diagnosis.

\subsubsection{Evaluation 1: Comparison study with state-of-the-art 3D convolutional neural networks}\hfil

\par
\begin{figure}[h]
    \centering
    \includegraphics[width=0.45\textwidth]{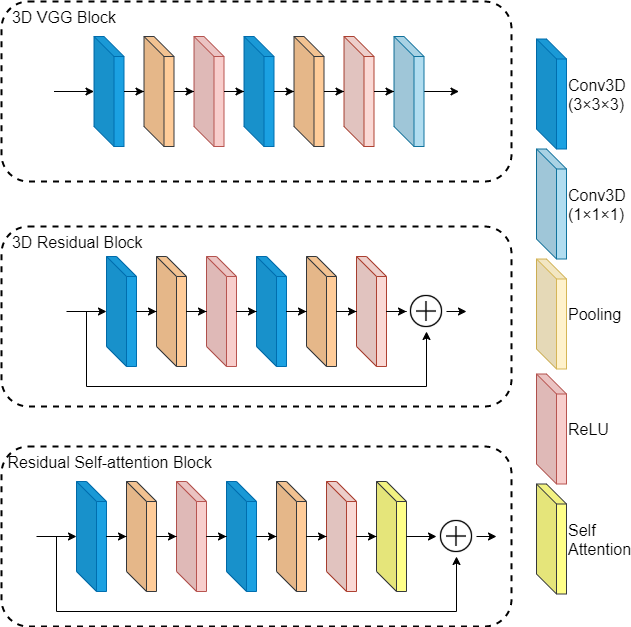}
    \caption{The structure of 3D-VGG Block, 3D-ResNet Block and 3D-ResAttNet Block.}
    \label{FIG:4}
\end{figure}

 We have performed comparison study with most commonly used 3D convolutional neural networks including 3D-VGGNet, 3D-ResNet under two conditions: with and without self-attention mechanism in 18 and 34 layers. The structures of 3D-VGG Block, 3D-ResNet Block and 3D-ResAttNet Block are shown in {Fig.~\ref{FIG:4}}.

\par
\begin{figure}[h]
    \centering
    \includegraphics[width=0.45\textwidth]{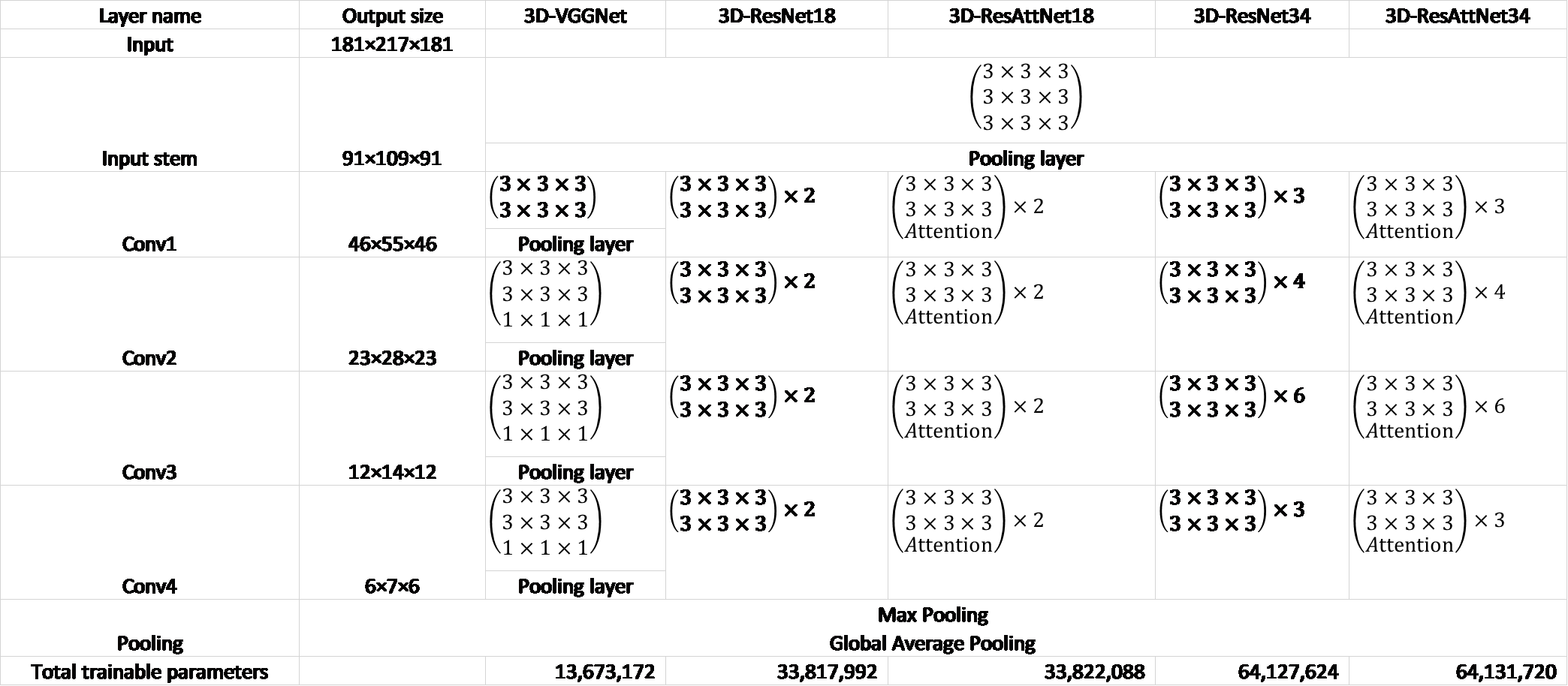}
    \caption{Overall architecture of 3D CNNs models, including 3D-VGGNet, 3D-ResNet 18, 3D-ResAttNet 18, , 3D-ResNet 34 and , 3D-ResAttNet 34.}
    \label{FIG:5}
\end{figure}

 Each Conv3D layer consists of 3 consistent operations: 3D convolution, batch normalization 3D and RELU. The 14 layers and 34 layers network contain 8 and 14 3D Resnet block and 3D-ResAttNet block, respectively. A $3\times3\times3$ 3D convolution is 3 times more expensive than 2D version in terms of computational cost. In order to reduce the computational cost, we have replaced the $7\times7\times7$ convolution in the 3D Conv block with three conservative $3\times3\times3$ convolutions. The detailed configuration is shown in {Fig.~\ref{FIG:5}}.

\par In this evaluation, we have trained our model using ADNI-1 dataset and performed a 5 fold cross-validation. The dataset is randomly split into 5 groups where 4 groups (80\% of the dataset) are used for training and the rest are used for testing each time. The experimental results for classification performance are the average of the accuracies on the testing set across all folds, along with Standard deviation (Std). p-value are also used to evaluate the statistical significance.
To optimize model parameters, Adam, a stochastic optimization algorithm, with a batch size of 8 samples, has been used for optimization to train the proposed network \cite{kingma_adam_2017}. We firstly set the learning rate (LR) as $1\times 10^{-4}$. The LR is decreased to $1\times 10^{-6}$ with increased iterations. CrossEntropy has been selected as the loss function for this task \cite{de2005tutorial}).  
\par All the experiments have been implemented based on PyTorch and executed on a server with an Intel(R) Xeon(R) CPU E5-2650, NVIDIA 2080TI and 64 GB memory.

\subsubsection{Evaluation 2: Evaluation on generalizability of the proposed model using two independent datasets}\hfil

\par To investigate generalizability and reproducibility of our proposed model, we have conducted two groups of experimental evaluation as follows: 
\begin{itemize}
\item We have first built our model based on ADNI-1 and evaluated it using two independent datasets (ADNI-2 and ADNI-3 respectively). 
 \item Then we reversed the training and testing datasets where we have trained the model using ADNI-2, and evaluated it on ADNI-1 and ADNI3 respectively. 
 \end{itemize}
In this evaluation, we only performed AD vs. NC classification task due to insufficient pMCI and sMCI samples obtained from ADNI-2 and ADNI-3.

\subsubsection{Evaluation 3: Comparison study with other existing machine learning/deep learning methods for AD diagnosis}\hfil

\par For indirect evaluation, we have selected most recent and state-of-the-art machine learning methods reported in the literature for indirect comparison using baseline sMRI data from ADNI \cite{baron_vivo_2001,fan_compare:_2006,kloppel_automatic_2008,koikkalainen_multi-template_2011,liu_relationship_2016,moller_alzheimer_2016,salvatore_magnetic_2015}.


\subsection{Result and discussion}

\subsubsection{Results from evaluation 1} \hfil

\par
\begin{figure}[h]
    \centering
    \includegraphics[width=0.5\textwidth]{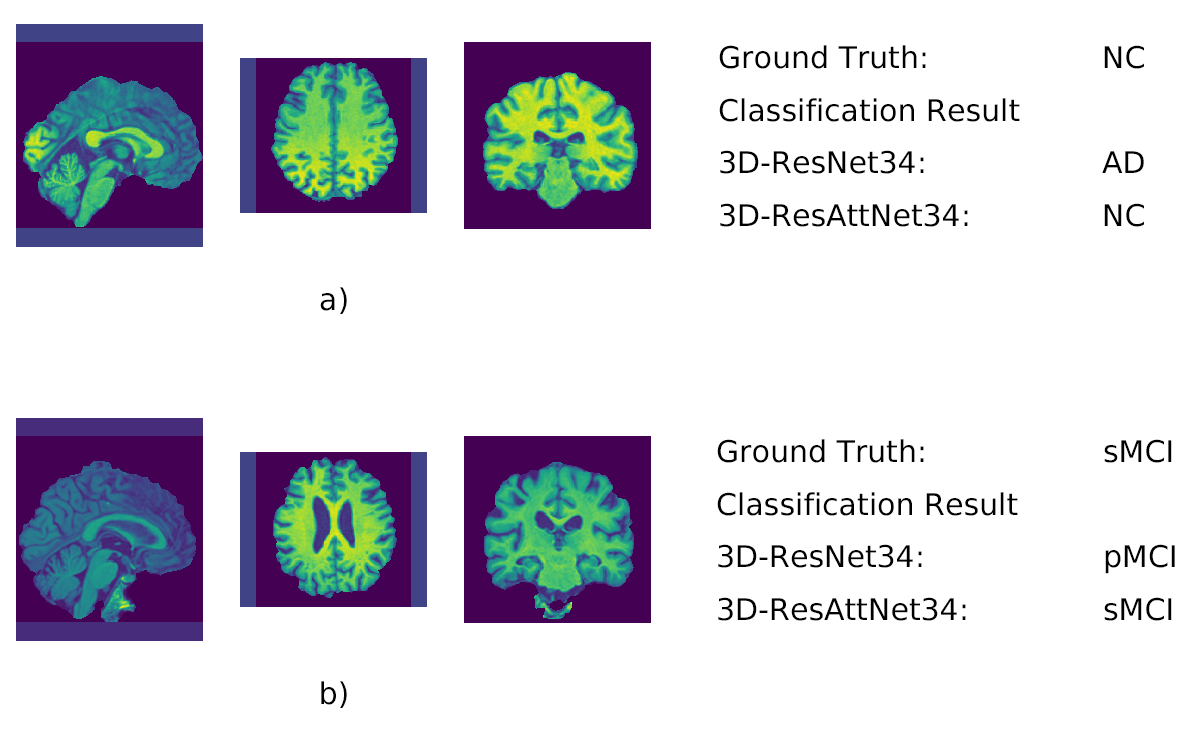}
    \caption{Examples of classification results on a) the AD vs. ND classification task. The result shows the proposed method classified the image into the right category same as the ground truth (NC) while 3D ResNet34 classified into a wrong catagory (AD);  b) the sMCI VS. pMCI task. The result shows the proposed method classified it into the right category same as the ground truth while 3D ResNet34 classified into a wrong category (pMCI) }
    \label{FIG:10}
\end{figure}

 As introduced previously, we have added the attention mechanism in the Resnet block. In this group of experiments, we have compared the models including 3D-VGGNet and 3D-ResNet models with and without attention layer. The results are presented in {Table~\ref{table:2}}. The classification performance of models with attention layer are significantly higher than models without it, especially on pMCI vs. sMCI classification. Our proposed model (3D-ResAttNet34) shows the best performance in all experiments. {Fig.~\ref{FIG:10}} shows the examples of classification results for two classification tasks: AD vs. NC and pMCI vs. sMCI tasks.  {Fig.~\ref{FIG:10}} a) shows an example on AD vs. NC classification where the classification result using our proposed model 3D-ResAttNet34 with attention layer classifies the image into a normal category (i.e. NC) while the result from 3D-ResNet34 classifies the image into disease category (i.e. AD). Similarly, {Fig.~\ref{FIG:10}} b) shows an example on pMCI vs. sMCI classification.  It indicates that our model correctly identifies the images into the right categories.

\begin{table}[h]
\caption{Results of classification for AD vs. NC and pMCI vs. sMCI}\label{table:2}
\centering
\resizebox{0.45\textwidth}{!}{%
\begin{tabular}{cllll}
\hline
\multirow{2}{*}{Model} & \multicolumn{4}{c}{AD vs. NC   classification}                \\ \cline{2-5} 
                       & ACC $\pm$ Std     & SEN $\pm$ Std     & SPE $\pm$ Std     & AUC $\pm$ Std     \\ \hline
3D-VGGNet              & 0.807 $\pm$ 0.046 & 0.798 $\pm$ 0.049 & 0.829 $\pm$ 0.036 & 0.890 $\pm$ 0.036 \\
3D-ResNet18            & 0.851 $\pm$ 0.102 & 0.849 $\pm$ 0.103 & 0.855 $\pm$ 0.102 & 0.920 $\pm$ 0.076 \\
3D-ResAttNet18         & 0.860 $\pm$ 0.088 & 0.829 $\pm$ 0.119 & 0.903 $\pm$ 0.052 & 0.975 $\pm$ 0.023 \\
3D-ResNet34            & 0.882 $\pm$ 0.147 & 0.890 $\pm$ 0.141 & 0.883 $\pm$ 0.148 & 0.929 $\pm$ 0.089 \\
3D-ResAttNet34         & 0.913 $\pm$ 0.012 & 0.910 $\pm$ 0.014 & 0.919 $\pm$ 0.009 & 0.984 $\pm$ 0.009 \\ \hline
\multirow{2}{*}{Model} & \multicolumn{4}{c}{pMCI vs. sMCI classification}              \\ \cline{2-5} 
                       & ACC $\pm$ Std     & SEN $\pm$ Std     & SPE $\pm$ Std     & AUC $\pm$ Std     \\ \hline
3D-VGGNet              & 0.758 $\pm$ 0.059 & 0.751 $\pm$ 0.083 & 0.735 $\pm$ 0.060 & 0.856 $\pm$ 0.056 \\
3D-ResNet18            & 0.777 $\pm$ 0.079 & 0.775 $\pm$ 0.092 & 0.753 $\pm$ 0.075 & 0.890 $\pm$ 0.034 \\
3D-ResAttNet18         & 0.799 $\pm$ 0.071 & 0.810 $\pm$ 0.093 & 0.775 $\pm$ 0.071 & 0.926 $\pm$ 0.053 \\
3D-ResNet34            & 0.807 $\pm$ 0.047 & 0.826 $\pm$ 0.054 & 0.798 $\pm$ 0.055 & 0.954 $\pm$ 0.033 \\
3D-ResAttNet34         & 0.821 $\pm$ 0.092 & 0.812 $\pm$ 0.101 & 0.809 $\pm$ 0.097 & 0.920 $\pm$ 0.047 \\ \hline
\end{tabular}%
}
\end{table}


\par
\begin{figure}[h]
    \centering
    \includegraphics[width=0.5\textwidth]{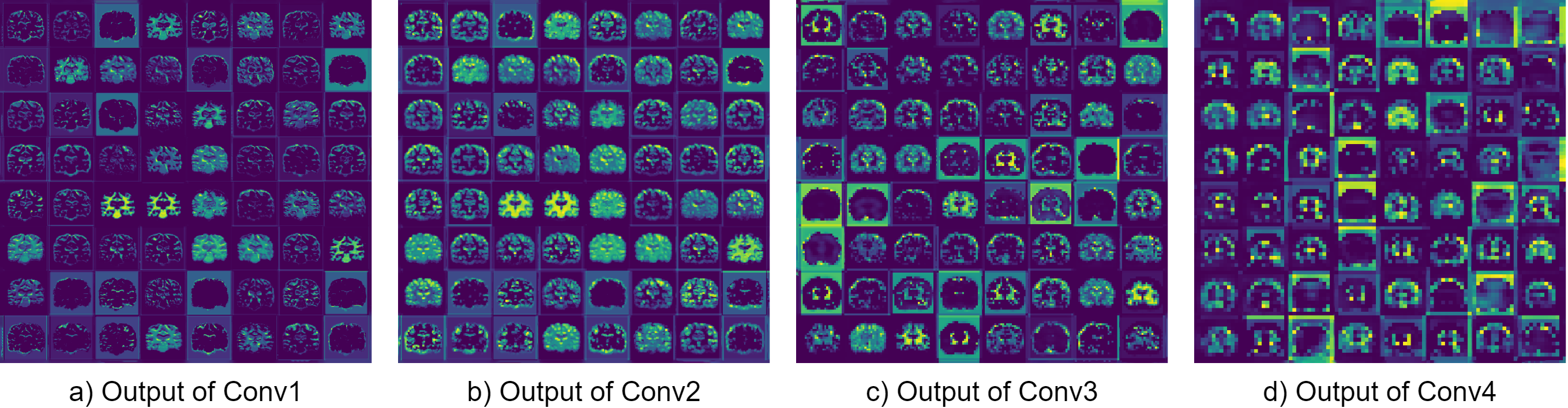}
    \caption{Visualization results of selected convolutional layer feature maps. From left to right: first, second, third and fourth convolutional block.}
    \label{FIG:9}
\end{figure}

 For the explainable model evaluation, 3D Grad-CAM has been used to explain the model for Alzheimer's Disease Diagnosis. We have only applied the 3D Grad-CAM to 3D-ResAttNet34 with the best classification performance. The heat-map is created to show how the network learns the importance of the areas. As described earlier, the 3D Grad-CAM can be used on an arbitrary layer. {Fig.~\ref{FIG:9}} shows feature visualizations of each convolution block from our proposed model (the top 64 activation maps are selected here). As more convolutions are processed, the resolution of the feature map also gradually decreases. In the first two convolution layers, the results have higher resolutions, which provide more details information. However, because they respond more to the corners, edge, texture and color conjunctions, more edges are highlighted. In the third and fourth convolution layers, the feature maps look like binary patterns where global and semantic information can be extracted. Some significant variation in lateral ventricle and hippocampus areas are highlighted.

\begin{figure}[h]
    \centering
    \includegraphics[width=0.45\textwidth]{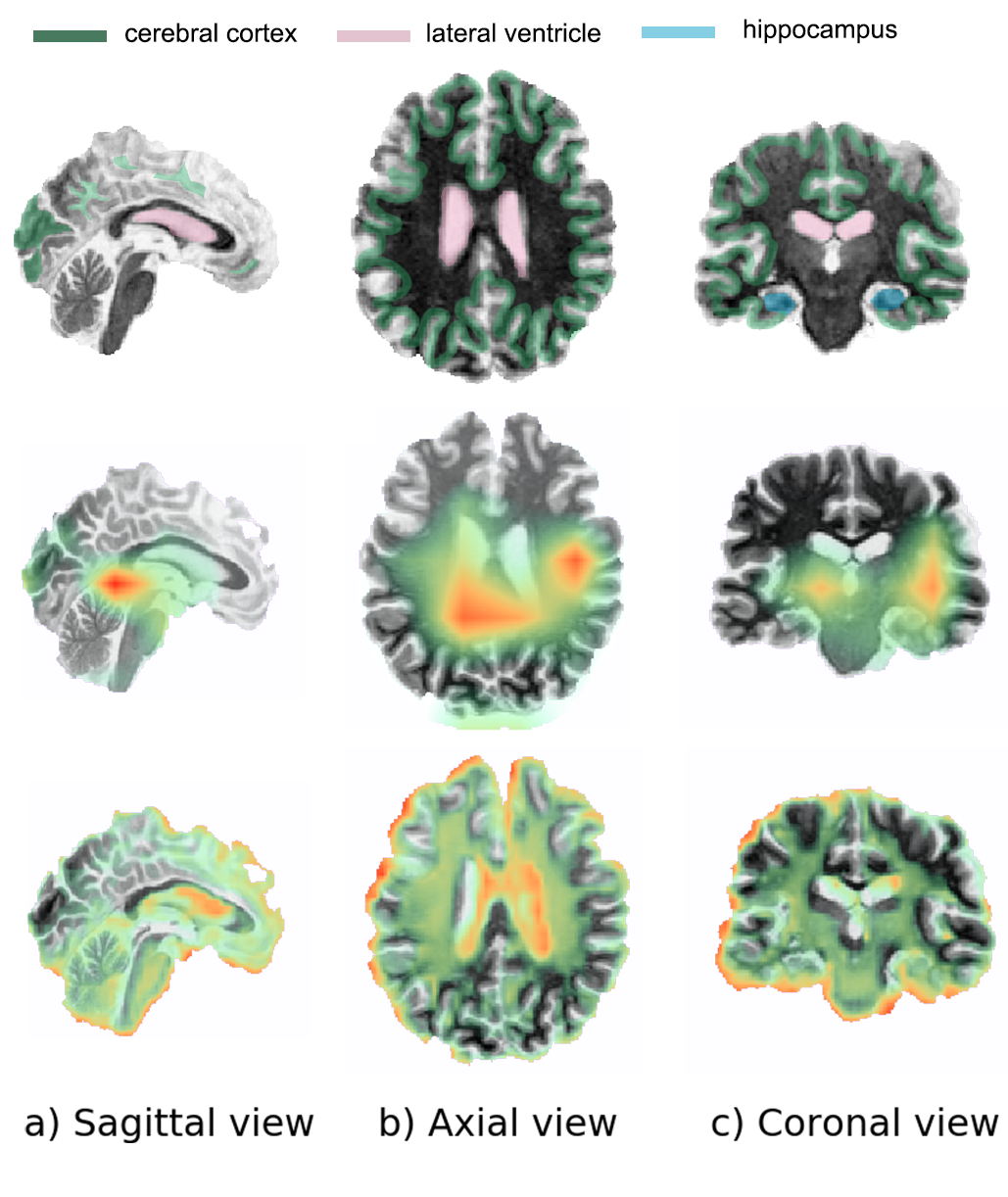}
    \caption{Sagittal, Axial and coronal view of the brain MRI and the visual explanation heatmaps.  The first row shows the highlighted cerebral cortex, lateral ventricle, and hippocampus areas in sMRI images. The second row shows the visualization by applying the Grad-Cam to the fourth convolutional layer.  The third row shows the visualization by applying the Grad-Cam to the third layer.}
    \label{FIG:8}
\end{figure}

The attention heatmap of the Grad-CAM on 3D-ResAttNet34 result is presented in {Fig.~\ref{FIG:8}}. For comparison, the hippocampus, lateral ventricle and cerebral cortex areas on the input sMRI image in the first row are labeled to show the important areas for Alzheimer’s disease diagnosis. In the second row, we have applied the activation mapping heat-map to the last convolutional layer (i.e the fourth layer in this case). The heatmap is blurry because the last convolutional layer of 3D-ResAttNet34 is only of size $6\times7\times6$. The heat map tends to show global information. To obtain a higher resolution and more detailed 3D class activation mapping heat-map, we have applied the 3D Grad-CAM to the lower convolutional layer (the third layer), as shown in the third row of {Fig.~\ref{FIG:8}}. It is of size $46\times55\times46$ heat-map and thus provides more detail information. It identifies and highlights the hippocampus, lateral ventricle and most parts of the cortex as important areas, which matches the human expert's approach \cite{mu_adult_2011,ott_brain_2010}. However, as mentioned in \cite{zeiler2014visualizing}, the lower layer in deep CNN models responds more to corners and edge/color conjunctions. Therefore, edges are highlighted as well.

\subsubsection{Results for evaluation 2} \hfil

\par The evaluation result for generalizability of the proposed model are summarized in {Table~\ref{table:3}}.
\par For the first group of experiments, we have trained model using ADNI-1 dataset and evaluated it on ADNI-2 and ADNI-3 respectively. Comparing to our model based on ADNI-1 data,  the accuracy on ADNI-2 is slightly decreased by 0.004 and the accuracy on ADNI-3 is slightly decreased by 0.021. The AUC is slightly dropped by 0.032 and 0.095 respectively. However, ACC, SEN, and SPE of our model remain high, which are statistically significant (i.e., p-values \textless{}0.05). This shows the good generalizabiity of our proposed model.
\par For the second group of experiments, we have reversed the training and testing datasets to train the model using ADNI-2 and evaluate it on ADNI-1 and ADNI-3 respectively. The accuracy of our proposed model on ADNI-2 reaches 0.956. When testing on ADNI-1 and ADNI-3, the accuracies are 0.933 and 0.917 respectively, with slight decreases by 0.23 on ADNI-1 and 0.39 on ADNI-3 respectively. The ACC, SEN, and SPE of our proposed model based on ADNI-2 are higher than the ones of the model using ADNI-1, which are statistically significant (i.e., p-values \textless{}0.05).  The main reason is because ADNI-1 and ADNI-2 dataset are captured from distinct phases of the ADNI project, which have different signal to noise ratios (SNR). sMRI images from ADNI-1 are scanned using 1.5T scanners, while MR images from ADNI-2 are scanned using 3T scanners. The 3T scanner has twice sensitive compared to 1.5T which can generate clearer and higher quality image.
\par Based on these experiments, it has demonstrated that our proposed approach has good generalizability and reproducibility for AD diagnosis.
\begin{table}[h]
\caption{The model performance on independent datasets (ADNI-2 and ADNI-3) }\label{table:3}
\Huge
\centering
\resizebox{0.5\textwidth}{!}{%
\begin{tabular}{cllllll}
\hline
\multirow{2}{*}{Model}          & \multicolumn{1}{c}{\multirow{2}{*}{Train}} & \multicolumn{1}{c}{\multirow{2}{*}{Evaluation}} & \multicolumn{4}{c}{AD vs. NC   classification}                \\ \cline{4-7} 
                                & \multicolumn{1}{c}{}                               & \multicolumn{1}{c}{}                                  & ACC $\pm$ Std     & SEN $\pm$ Std     & SPE $\pm$ Std     & AUC $\pm$ Std     \\ \hline
\multirow{6}{*}{\begin{tabular}[c]{@{}c@{}}Proposed\\ 3D-ResAttNet34\end{tabular}} & ADNI-1                                             & ADNI-1                                                & 0.913 $\pm$ 0.012 & 0.910 $\pm$ 0.014 & 0.919 $\pm$ 0.009 & 0.984 $\pm$ 0.009 \\
                                & ADNI-1                                             & ADNI-2                                                & 0.909 $\pm$ 0.019 & 0.895 $\pm$ 0.026 & 0.924 $\pm$ 0.014 & 0.952 $\pm$ 0.026 \\
                                & ADNI-1                                             & ADNI-3                                                & 0.892 $\pm$ 0.034 & 0.788 $\pm$ 0.060 & 0.780 $\pm$ 0.076 & 0.889 $\pm$ 0.034 \\ \cline{2-7} 
                                & ADNI-2                                             & ADNI-2                                                & 0.956 $\pm$ 0.089 & 0.950 $\pm$ 0.100 & 0.950 $\pm$ 0.100 & 0.978 $\pm$ 0.044 \\
                                & ADNI-2                                             & ADNI-1                                                & 0.933 $\pm$ 0.133 & 0.933 $\pm$ 0.133 & 0.930 $\pm$ 0.140 & 0.967 $\pm$ 0.067 \\
                                & ADNI-2                                             & ADNI-3                                                & 0.917 $\pm$ 0.019 & 0.885 $\pm$ 0.068 & 0.821 $\pm$ 0.046 & 0.930 $\pm$ 0.023 \\ \hline
\end{tabular}%
}
\end{table}


\subsubsection{Results for evaluation 3} \hfil

\par It is unfair to perform the direct comparison between different methods due to the use of different datasets and also the clinical definition of pMCI/sMCI.  In this case, we have only indirectly compared our model with six state-of-the-art machine learning based methods \cite{liu_early_2014, suk_hierarchical_2014,aderghal_classification_2017,liu_landmark-based_2018,shi_multimodal_2018, lian_hierarchical_2018}. The results are shown in {Table~\ref{table:4}}. 
There are several important observations including: 1) for  the challenging task of MCI conversion prediction, the proposed 3D-ResAttNet outperforms other existing approaches; 2) for AD vs, NC classification, our proposed method has competitive performance, comparing to those methods using MRI only; 3) In terms of data size, our proposed method has been evaluated on a large number of subjects and cross-validated on two independent datasets from ADNI-2 and ADNI-3 respectively, which demonstrates a fair and independent evaluation and a good generalizability of our model. In addition, compared with the traditional region- and voxel-level pattern analysis methods, our proposed method takes the whole MRI image as input and automatically extracts high dimensional and nonlinear features, which leads to better classification performance for AD diagnosis.

\begin{table}[h]
\caption{Comparative performance of the classifier vs. six competitors on ADNI dataset.}\label{table:4}
\Huge
\centering
\resizebox{0.5\textwidth}{!}{%
\begin{tabular}{llllllllll}
\hline
\multirow{2}{*}{References} & \multicolumn{1}{c}{\multirow{2}{*}{Modality}} & \multicolumn{1}{c}{\multirow{2}{*}{Subject}} & \multirow{2}{*}{Method}     & \multicolumn{3}{l}{AD vs. NC classification} & \multicolumn{3}{l}{pMCI vs. sMCI classification} \\ \cline{5-10} 
                            & \multicolumn{1}{c}{}                          & \multicolumn{1}{c}{}                         &                             & ACC           & SEN           & SPE          & ACC            & SEN            & SPE            \\ \hline
(Liu et al., 2014)          & PET/MRI                                      & 65AD+169MCI+77NC                     & Stacked auto-encoder        & 0.88          & 0.89          & 0.87         & 0.77           & 0.74           & 0.78           \\
(Suk et al., 2014)          & PET/MRI                                      & 93AD+204MCI+101NC                            & Deep Boltzmann machine      & 0.95          & 0.95          & 0.95         & 0.76           & 0.48           & 0.95           \\
(Aderghal et al., 2017a)    & MRI                                           & 188AD+399MCI+228NC                           & 2D-CNN                      & 0.91          & 0.94          & 0.89         & 0.66           & 0.66           & 0.65           \\
(Liu et al., 2018)          & MRI                                           & 199AD+393MCI+229NC                  & Landmark detection + 3D CNN & 0.91          & 0.88          & 0.94         & 0.77           & 0.42           & 0.82           \\
(Shi et al., 2018)          & MRI                                           & 41AD+99MCI+52NC                      & Deep polynomial network     & 0.95          & 0.94          & 0.96         & 0.75           & 0.63           & 0.85           \\
(Lian et al., 2018)         & MRI                                           & 358AD+670MCI+429NC                  & Hierarchical-CNN            & 0.90          & 0.82          & 0.97         & 0.81           & 0.53           & 0.85           \\
Our 3D-ResAttNet on ADNI-1  & MRI                                           & 200AD+404MCI+231NC                  & 3D-CNN                      & 0.91          & 0.91          & 0.92         & 0.82           & 0.81           & 0.81           \\ \hline
\end{tabular}%
}
\end{table}

\section{Conclusion}

\par Inspired by the attention mechanism and residual learning, we have proposed an end-to-end framework based on 3D Residual Self-Attention Network (3D ResAttNet) for early efficient diagnosis of AD diseases at two levels (i.e., AD vs. NC and pMCI vs. sMCI) from sMRI scans.  The proposed method combines residual learning with self-attention mechanism, which can fully exploit both global and local information and avoid the information loss. Meanwhile, to understand inside our model and how our model reach decisions, we have also applied the 3D Grad-CAM method to identify and visualize those important areas contributing to our model decisions. To evaluate our model performance, we have compared the proposed model with most commonly used 3D convolutional neural networks including 3D-VGGNet, 3D-ResNet. The results show that our proposed model with attention layer (3D ResAttNet)  outperforms the existing models.  To evaluate generalization of the proposed model, we have also conducted thorough experiments under different cross-validation strategies using ADNI datasets (ADNI-1, ADNI-2 and ADNI-3): 1) building the proposed model based on ADNI-1 and then validating it on ADNI-2 and ADNI-3 respectively; 2) building the model based on ADNI-2 and then validating it on ADNI-1 and ADNI-3 respectively.  The results show that our proposed model has a good generalizability in all cases. Moreover, the explainable mechanism in our approach is able to identify and highlight the contribution of the important brain parts (e.g., hippocampus, lateral ventricle and most parts of the cortex) for transparent decisions. The future work will focus on continuous improvement of model performance and generalizability using more independent datasets.

\section*{Acknowledgment}

The work reported in this paper has formed part of the project by Royal Society - Academy of Medical Sciences Newton Advanced Fellowship (NAF$\backslash$R1$\backslash$180371).

\ifCLASSOPTIONcaptionsoff
  \newpage
\fi



\bibliographystyle{IEEEtran}
\bibliography{mri,test}
%




%

\begin{IEEEbiography}[{\includegraphics[width=1in,height=1.25in,clip,keepaspectratio]{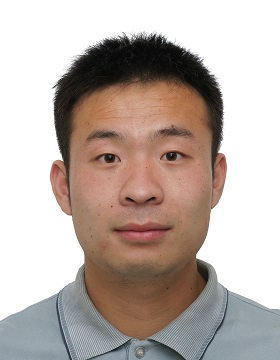}}]{Xin Zhang}
Xin Zhang is associate researcher in Manchester Metropolitan University (MMU), he received the B.S degree from The PLA Academy of Communication and Commanding, China, in 2009 and Ph.D. degree in Cartography and Geographic Information System from Beijing Normal University(BNU), China, in 2014. His current research interests include remote sensing image processing and deep learning.
\end{IEEEbiography}
\vskip -0.2in

\begin{IEEEbiography}[{\includegraphics[width=1in,height=1.25in,clip,keepaspectratio]{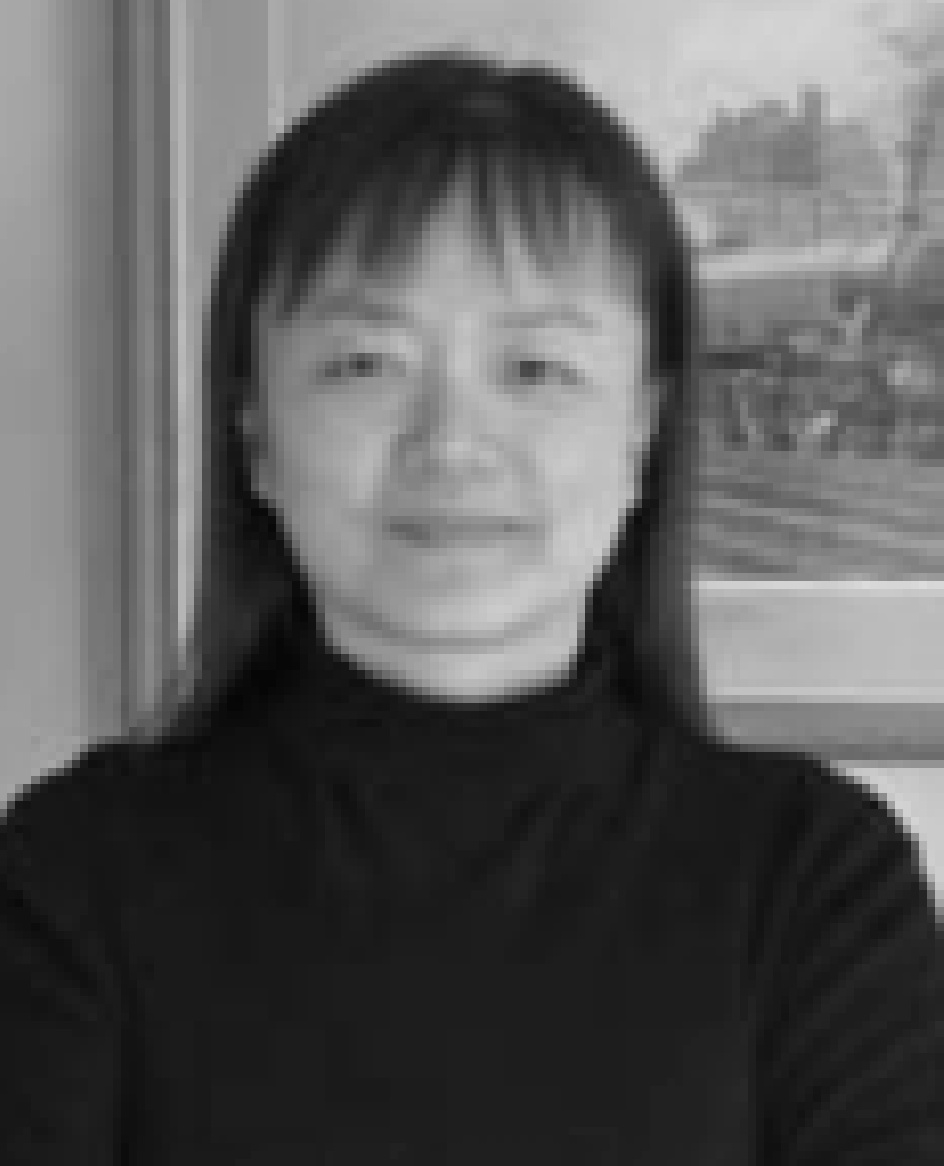}}]{Liangxiu~Han} received the Ph.D. degree in computer science from Fudan University, Shanghai, China, in 2002. She is currently a Professor of computer science with the School of Computing, Mathematics, and Digital Technology, Manchester Metropolitan University.  Her research areas mainly lie in the development of novel big data analytics and development of novel intelligent architectures that facilitates big data analytics (e.g., parallel and distributed computing, Cloud/Service-oriented computing/data intensive computing) as well as applications in different domains using various large datasets (biomedical images, environmental sensor, network traffic data, web documents, etc.). She is currently a Principal Investigator or Co-PI on a number of research projects in the research areas mentioned above.
\end{IEEEbiography}
\vskip -0.2in

\begin{IEEEbiography}[{\includegraphics[width=1in,height=1.25in,clip,keepaspectratio]{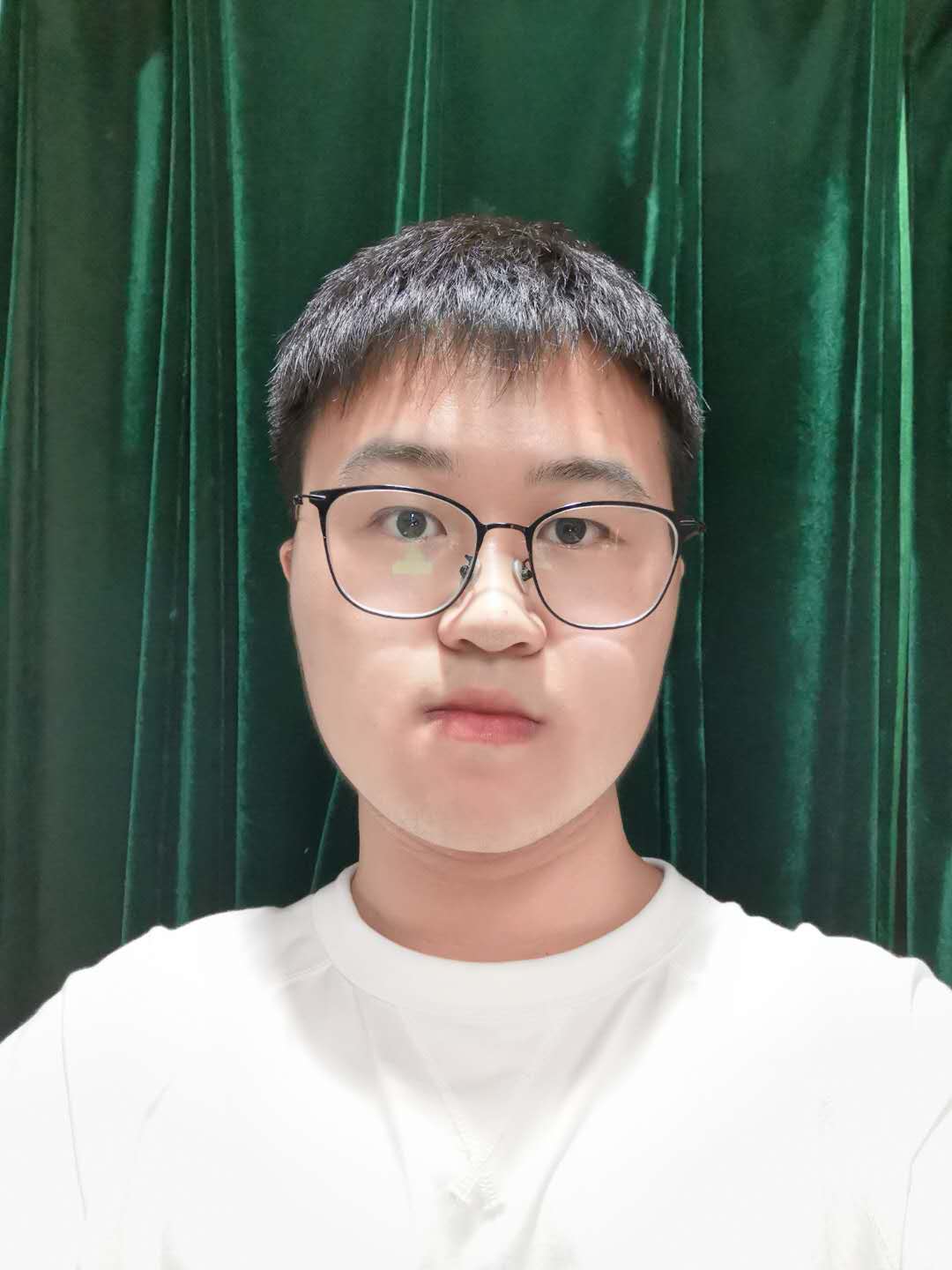}}]{Wenyong Zhu}
Wenyong Zhu received the B.S degree from Nanjing University of Aeronautics and Astronautics (NUAA), China, in 2019. And he is studying for a master's degree in NUAA. His current research interests include machine learning and medical image classification.
\end{IEEEbiography}
\vskip -0.2in

\begin{IEEEbiography}[{\includegraphics[width=1in,height=1.25in,clip,keepaspectratio]{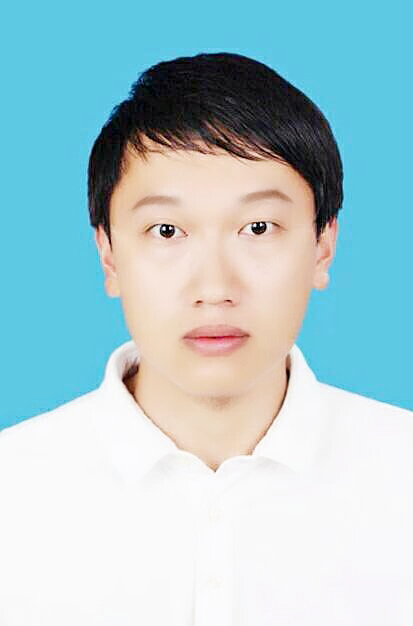}}]{Sun Liang}
Liang Sun received the B.S degree from Shandong University of Science and Technology, China, in 2014, and Ph.D. degree in Computer Science and Technology from Nanjing University of Aeronautics and Astronautics (NUAA), China, in 2020. His current research interests include machine learning and medical image segmentation.
\end{IEEEbiography}

\vskip -0.2in

\begin{IEEEbiography}[{\includegraphics[width=1in,height=1.25in,clip,keepaspectratio]{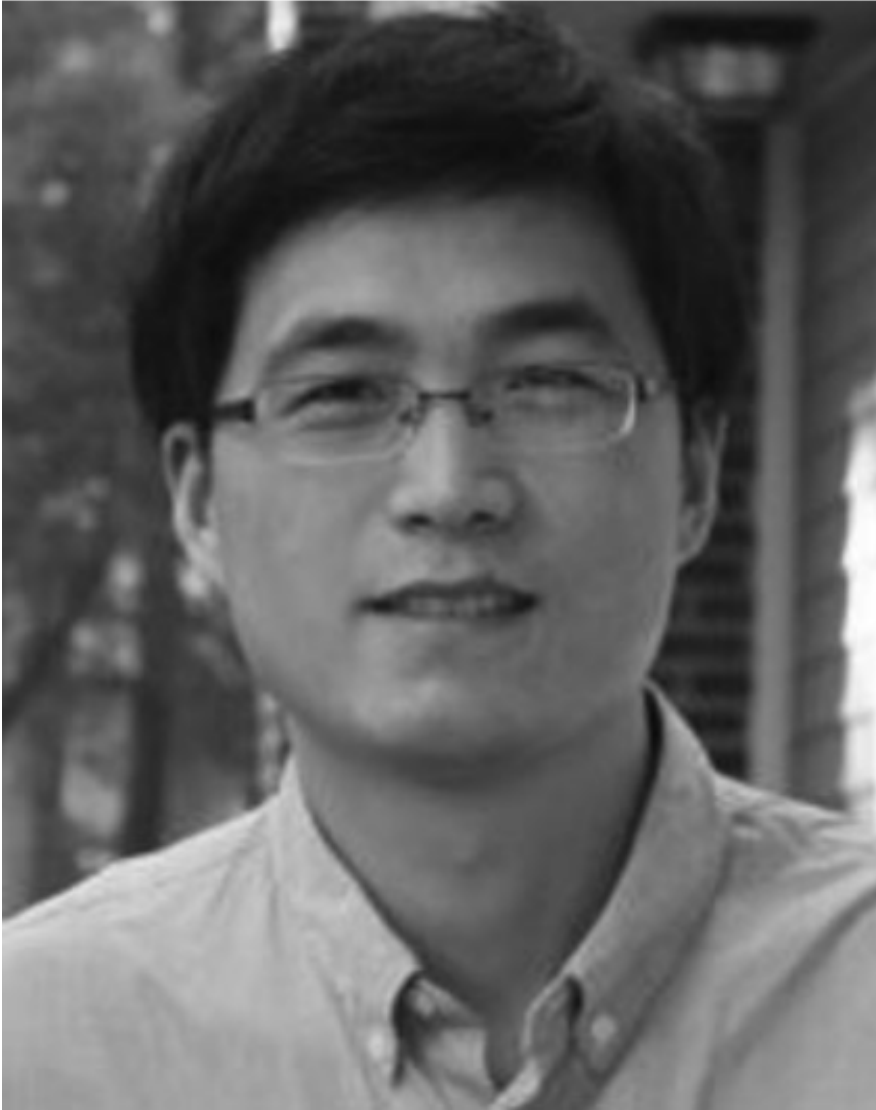}}]{Daoqiang~Zhang}
	received the B.Sc. and Ph.D. degrees in computer science from Nanjing University of Aeronautics and Astronautics, Nanjing, China, in 1999 and 2004, respectively. He is currently a Professor in the Department of Computer Science and Engineering, Nanjing University of Aeronautics and Astronautics. His current research interests include machine learning, pattern recognition, and biomedical image analysis. In these areas, he has authored or coauthored more than 100 technical papers in the refereed international journals and conference proceedings. 
\end{IEEEbiography}
\vskip 4in



\end{document}